# Preparation of hierarchical C@MoS$_2$@C sandwiched hollow spheres for Lithium ion batteries

Zhenyou Li*[1,2], Alexander Ottmann[1], Ting Zhang[2], Qing Sun[1], Hans-Peter Meyer[4], Yana Vaynzof[1,3], Junhui Xiang[§2], Rüdiger Klingeler[#1,3]

[1]*Kirchhoff Institute of Physics, Heidelberg University, INF 227, 69120 Heidelberg, Germany*

[2]*College of Materials Science and Opto-Electronic Technology, University of Chinese Academy of Sciences, Yuquan Road 19A, Beijing, 100049 China.*

[3]*Centre for Advanced Materials (CAM), Heidelberg University, INF 225, 69120 Heidelberg, Germany*

[4]*Institute of Earth Sciences, Heidelberg University, INF 236, D-69120 Heidelberg*

*Author Information*

[§]Email: xiangjh@ucas.ac.cn

[#]Email: klingeler@kip.uni-heidelberg.de

*Corresponding Authors*

*Email: zhenyou.li@kip.uni-heidelberg.de

*Author Contributions*

The manuscript was written through contributions of all authors. All authors have given approval to the final version of the manuscript.

**Abstract**

Hierarchical C@MoS$_2$@C hollow spheres with the active MoS$_2$ nanosheets being sandwiched by carbon layers have been produced by means of a modified template method. The process applies polydopamine (PDA) layers which inhibit morphology change of the template thereby enforcing the hollow microsphere structure. In addition, PDA forms complexes with the Mo precursor, leading to an in-situ growth of MoS$_2$ on its surface and preventing the nanosheets from agglomeration. It also supplies the carbon that finally sandwiches the 100-150 nm thin MoS$_2$ spheres. The resulting hierarchically structured material provides a stable microstructure where carbon layers strongly linked to MoS$_2$ offer efficient pathways for electron and ion transfer, and concomitantly buffer the volume changes inevitably appearing during the charge-discharge process. Carbon-sandwiched MoS$_2$-based electrodes exhibit high specific capacity of approximately 900 mA h g$^{-1}$ after 50 cycles at 0.1 C, excellent cycling stability up to 200 cycles, and superior rate performance. The versatile synthesis method reported here offers a general route to design hollow sandwich structures with a variety of different active materials.

**Introduction**

Hierarchically structured nanomaterials, appropriately designed to combine functionalities mandatory for high capacitance and/or high power electrode materials, are a promising avenue towards improved lithium-ion batteries (LIB).[1, 2] In such materials, differently sized building blocks with different associated functionalities can be combined, not by simply mixing but by rationally arranging the components in order to precisely control the functionalities of the final product.[3] This particularly holds for conversion materials whose high theoretical capacity by far exceeds that of traditional intercalation materials since the conversion reactions are associated with multi-electron transfer per metal centre.[4] However, corresponding large volume changes upon charging/discharging raise thermodynamic and kinetic issues. Downsizing the active materials as well as hierarchical structuring either with or without carbon coating are effective strategies to overcome these problems as it enables buffering the strain caused by the volume changes, increases the contact area between active material and electrolyte, and shortens the diffusion distance of lithium ions.[5]

With respect to electrochemical energy storage, two dimensional layered $MoS_2$ has recently come into the research focus because it is not only an intercalation material, but can also serve as a conversion reaction electrode for LIB.[6-8] Based on the latter mechanism, deep discharge associated with storage of four Li-ions per formula unit is feasible, providing a favourable high theoretical capacity of ~670 mA h $g^{-1}$.[9-11] However, the abovementioned typical features of conversion reaction materials have been found to result in issues such as capacity fading and poor rate performance.[5, 12] Therefore, much effort has been devoted to materials design by producing various $MoS_2$ nanostructures including hollow spheres[13], tubes[14, 15], nanoboxes[9], nanoflowers[16], nanoflakes[17], etc., which indeed show to a certain extent improved electrochemical properties. However, advantageous high specific surface areas of these nanomaterials are accompanied by low tap densities and high chemical activities which induce, for example, serious agglomeration, large interparticle resistances, and unwanted side reactions. In consequence, $MoS_2$-based materials are still far from commercial use in LIB.[18]

Very recently, progress in design of $MoS_2$-based anode materials was achieved by means of hierarchical hollow particles which are found to at least partly resist the destruction of the initial structures upon cycling.[19-21] While carbon is usually added in order to fabricate the electrodes[22], recent findings suggest that making $MoS_2$/C composite,[23, 24] especially with large contact area between carbon and $MoS_2$,[3] improves the battery performance greatly. Inspired by this fact, we have developed a preparation method of C@$MoS_2$@C sandwich structures forming hierarchical hollow spheres. We report a modified template approach and show that the resulting hierarchically structured $MoS_2$/C-nanomaterial exhibits outstanding specific capacity, cycling stability and rate performance. The synthesis route presented here offers a facile and general way of designing hierarchical functional nanomaterials.

**Results and Discussion**

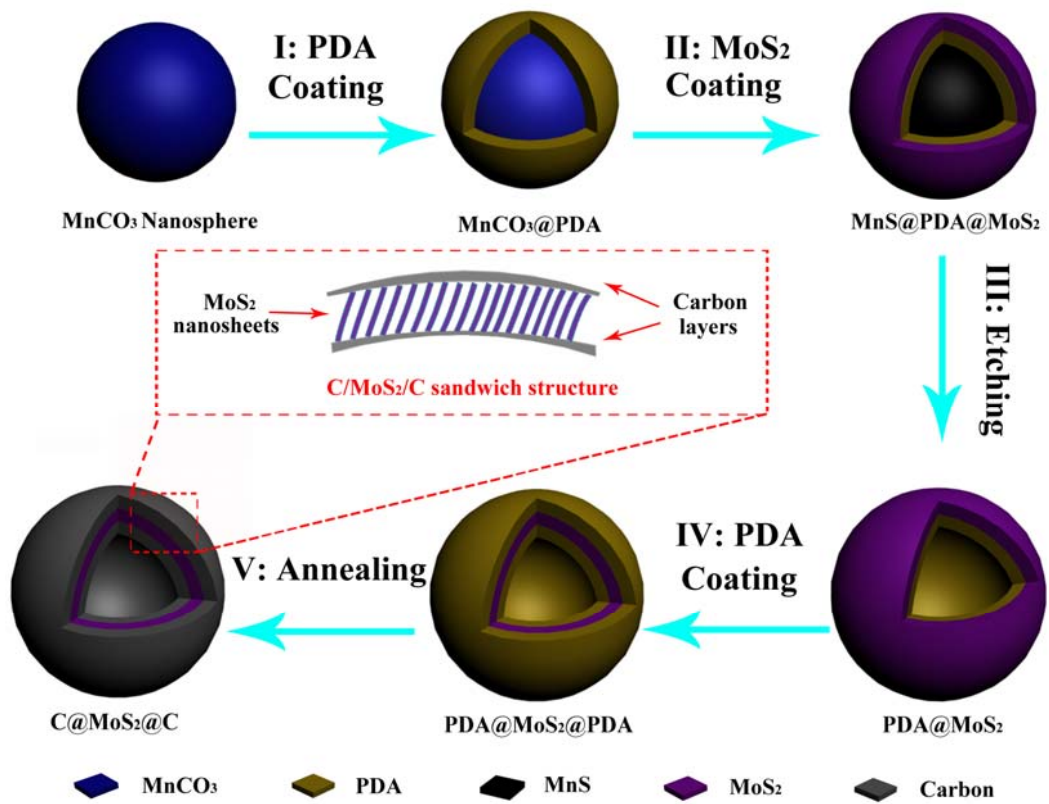

Fig. 1: Schematic illustration of the synthesis process and the hierarchical sandwich structure.

The synthesis of the hierarchical C@MoS$_2$@C sandwich structures by means of a modified template method is illustrated in Fig. 1. In general, sub-μm MnCO$_3$ spheres are used as template for the hollow structures which in a first step are coated by polydopamine (PDA) in order to shield the nanospheres against the following reaction steps. With the help of complexation between PDA and the Mo precursor, MoS$_2$ nanosheets grow in-situ on the spherical surface in a hydrothermal process. During that time, the MnCO$_3$ template is transformed to MnS because of the abundance of H$_2$S accompanying the formation of MoS$_2$. Under the protection of the PDA layer, the shape of the template still remains spherical. Comparatively, MoS$_2$/MnS hybrid cubes (Fig. S1 of the Supplement Information) are formed under the same experimental conditions but without PDA layer. After subsequent acid etching, and the addition of a second PDA layer, a final annealing step yields the desired C@MoS$_2$@C sandwiched hollow structure.

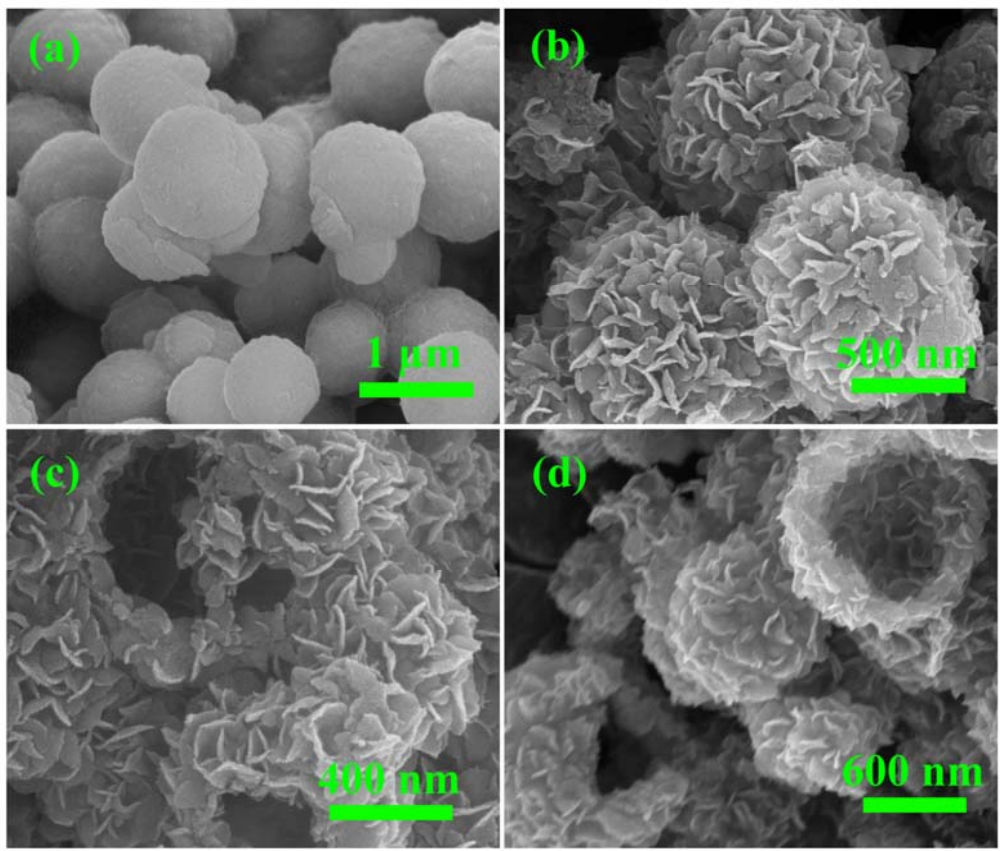

Fig. 2. SEM images of (a) MnCO$_3$@PDA, (b) MnS@PDA@MoS$_2$ spheres, (c) PDA@MoS$_2$ hollow spheres, (d) PDA@MoS$_2$@PDA hollow spheres.

The synthesis process has been confirmed by various characterization methods. The different intermediate products of the reaction have been investigated by means of XRD and SEM. As shown in Fig. S2 of the Supplement Information, the $MnCO_3$ template is of spherical shape with diameters in the range between 400 to 800 nm. The XRD pattern (Fig. S3 of the SI) confirms the rhombohedral $MnCO_3$ structure (JCPDS card #41-1472). Fig. 2a shows that the spherical shape is maintained after PDA coating, i.e. synthesis step I (Fig. 1). Since the PDA coating is essential to form the hierarchical structure, a control experiment was performed in order to show the stability of PDA under the hydrothermal conditions applied during the synthesis procedure. In this experiment, the PDA coated $MnCO_3$ spheres ($MnCO_3$@PDA) were hydrothermally treated at 200 °C for 24 h. After the same acid etching (2 M HCl solution for 24 h) and sintering (900 °C for 10 h under Ar atmosphere) which is applied to obtain the C@$MoS_2$@C sandwich structure, hollow carbon spheres were obtained as shown in Fig. S4 of the SI. The XRD pattern in Fig. S5 shows no $MnCO_3$ peaks but only a broad peak at around 2θ = 20-30°, which is characteristic for amorphous carbon.[25] The XPS spectrum in Fig. S6 confirms the presence of only C, O, and N in the PDA-derived carbon hollow spheres, which further indicates the presence of PDA layers. Due to the interaction between the residual phenolic hydroxyl groups on PDA layer and Mo precursors, $MoS_2$ nanosheets are grown in-situ on the surface of the PDA layers during a hydrothermal synthesis step (step II, Fig. 1). In this step, $MnCO_3$ reacts with $H_2S$ resulting from $(NH_4)_2MoS_4$, such that the inner template spheres transform to MnS. The SEM image of the resulting MnS@PDA@$MoS_2$ (Fig. 2b) shows microsphere-like outlines with about 1 μm in diameter. The surfaces of the microspheres are decorated with ~100 nm sized $MoS_2$ nanosheets in no particular order. After acid treatment with HCl (step III, Fig. 1), PDA@$MoS_2$ hollow structures are obtained (Fig. 2c). The inner diameter of these hollow spheres is around 700 nm, which is in good agreement with the size of the $MnCO_3$ nanospheres. This result reveals that with the help of the PDA layer the spherical morphology of the template is well preserved after the hydrothermal synthesis step, although it has been converted from $MnCO_3$ to MnS. Thereafter, the sandwich structure of the PDA@$MoS_2$@PDA hollow spheres is produced by another PDA coating (step IV,

Fig. 1). The microspheres of the resulting sandwich structure (Fig. 2d) show a thicker shell layer than the PDA@MoS$_2$ sample (Fig. 2c).

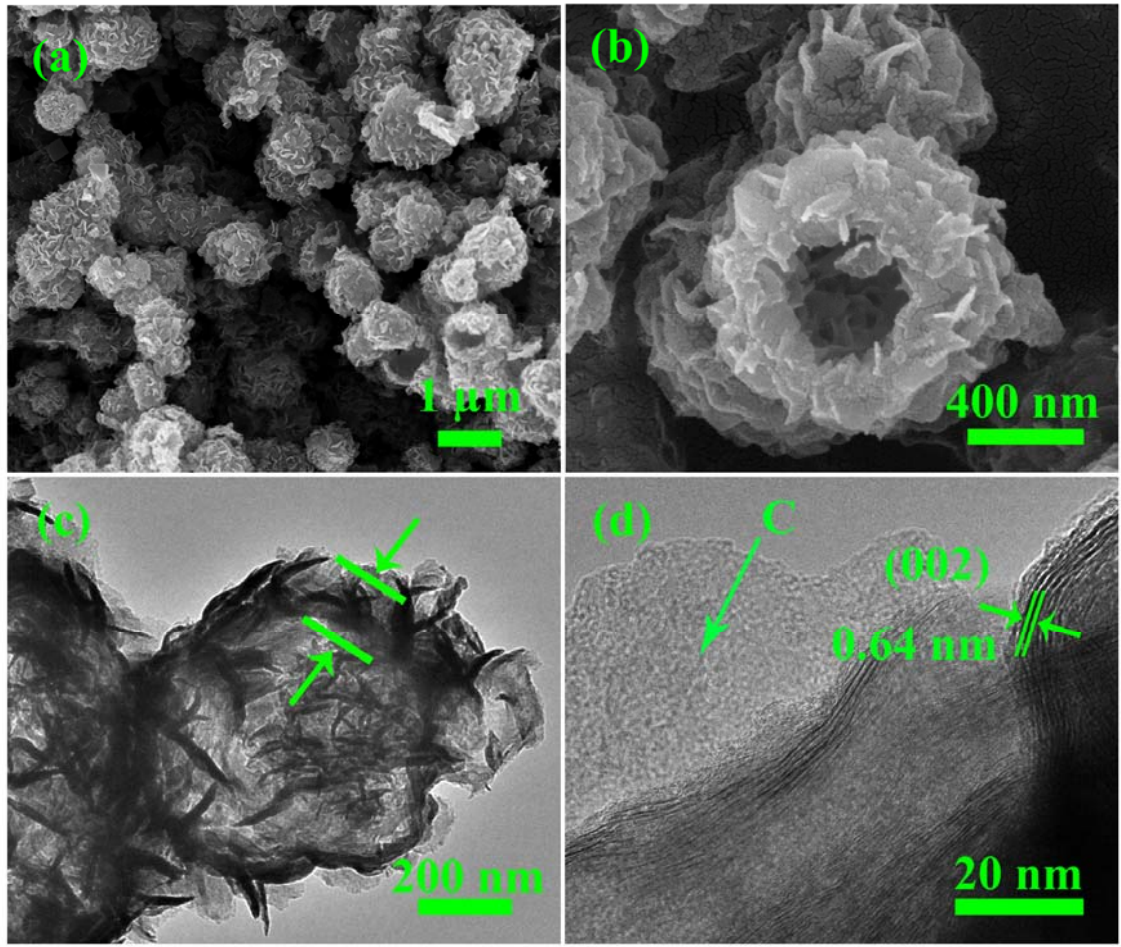

Fig. 3 SEM (a) ,(b) and TEM (c), (d) images of the C@MoS$_2$@C hollow spheres at different magnifications.

The final C@MoS$_2$@C hollow spheres are obtained by annealing the sandwiched PDA@MoS$_2$@PDA in Ar atomsphere. The hollow morphology and hierarchical structure of the C@MoS$_2$@C sample are seen in the SEM and TEM images presented in Fig. 3. The low magnification SEM picture (Fig. 3a) showing numerous microspheres confirm a rather narrow size distribution less than 1 μm. Some broken spheres reveal the hollow nature of the material (Fig. 3b). A typical hollow sphere of C@MoS$_2$@C, shown in the TEM image of Fig. 3c exhibits a loose MoS$_2$ array on the carbon shell with a diameter of about 800 nm and shell thickness of 100-150 nm which was marked by green arrows in Fig. 3c. TGA of the materials (Fig. S7) implies a carbon content of 25.9% in C@MoS$_2$@C, assuming that all carbon is combusted and MoS$_2$ is converted to MoO$_3$. Note, that the carbon content can be changed easily by controlling the thickness of the PDA layers. In virtue of the dispersive MoS$_2$ nanosheets on the carbon layer as well as the hollow character of the product, the C@MoS$_2$@C sample possesses a high specific surface area of 42.9 ± 0.5 m$^2$ g$^{-1}$ (Fig. S8a of the SI). This value is almost one order of magnitude larger than that of the pure MoS$_2$ assembly which consists of random MoS$_2$ nanosheet aggregates (Fig. S9 of the SI), and synthesized at the same hydrothermal conditions without templates. However, according to the pore distribution calculated by means of the Barrett–Joyner–Halenda (BJH) analysis (inset of Fig. S8a of the SI), the C@MoS$_2$@C sample does not show obvious mesoporous behaviour. It is worth noting that no gap between the carbon layer and the MoS$_2$ layer is observed, indicating tight binding of the two components. This strong interaction originates from abundant functional groups at the carbon layers derived from PDA. In fact, the phenolic hydroxyl groups of the PDA can chelate with Mo precursors,[26] offering active sites for MoS$_2$ growth and also preventing the nanosheets from agglomerating. The high resolution TEM image in Fig. 3d displays the fine structure of the sandwich design. We attribute the atomically disordered region which is marked by the green arrows to amorphous carbon layers. The legible lattice fringes in between are MoS$_2$ layers. The interlayer distance amounts to 0.64 nm which is larger than the (002) one of standard MoS$_2$ (0.62 nm)[27], indicating an expansion of the interlayer spacing.[28]

The crystal structure of the intermediates and of the final product are studied by XRD measurements as shown in Fig. 4a. The main diffraction peaks of the MoS$_2$@PDA@MnS sample (blue curve) are in good agreement with the standard MnS pattern (JCPDS No. 88-2223)[29] except for a small peak at 2θ ≈ 17.8° which corresponds to the (004) peak of MoS$_2$. As mentioned before, the MnS phase occurs due to the sulfuration of the MnCO$_3$ template. Accordingly, there are no MnS diffraction peaks in the diffraction pattern of the PDA@MoS$_2$@PDA hollow spheres (red curve), confirming the complete removal of the template. All peaks of this sample can be indexed in the hexagonal phase of MoS$_2$ (JCPDS No. 37-1492). The observed peak broadening indicates low crystallinity of the product. Comparatively, the C@MoS$_2$@C structure exhibits a more pronounced peak between 32° and 35° and a new peak around 40°, corresponding to the (100) and (103) plane of 2H-MoS$_2$, respectively. It is worth noting that the low angle diffraction patterns in Fig. 4b show obvious differences of the unsintered (PDA@MoS$_2$@PDA) and the sintered (C@MoS$_2$@C) samples. The shift of the (002) peak to lower angles and the appearance of the (004) peak at 17.9° in PDA@MoS$_2$@PDA can be attributed to the expansion of the interlayer distance due to the hydrothermal synthesis conditions. The dual peak feature of the unsintered sample in the small angle regime of 5 to 20° are merged to only one peak after sintering, which can be explained by a structural conversion to the thermodynamically stable MoS$_2$ phase in the final product.[30, 31] From the the (002) peak position of C@MoS$_2$@C, the interlayer distance of 0.64 nm is deduced which agrees well with the TEM data and confirms enlarged interlayer spacing compared as compared to the to the standard material (0.62 nm).[28]

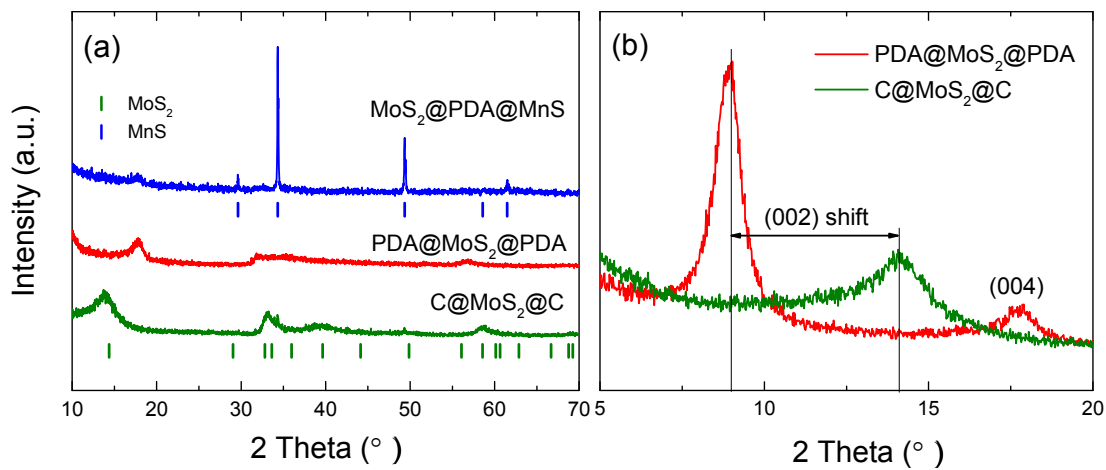

Fig. 4 (a) XRD patterns of the MoS$_2$@PDA@MnS, PDA@MoS$_2$@PDA hollow spheres, and C@MoS$_2$@C hollow spheres. The blue and green vertical ticks display the standard pattern of MoS$_2$ and MnS, respectively. (b) Small angle diffraction patterns of the sandwiched structures.

The surface chemical states and composition of the C@MoS$_2$@C hollow spheres were investigated by XPS (Fig. 5). In the survey scan, there are intensive peaks from S, Mo, C, N and O as well as weak peaks of Cu and Mn, the latter originating from the substrate and template residues. The high resolution scan of Mo3d shows two doublets: the doublet at 229.5 eV and 232.6 eV is attributed to the Mo 3d$_{5/2}$ and 3d$_{3/2}$ orbitals of MoS$_2$. The much weaker doublet at higher binding energy (marked in orange in Fig. 5b) indicates trace amounts of MoO$_x$ in the final product. The S2p region shows a single doublet at binding energies of 162.3 eV and 163.6 eV corresponding to the S2p$_{3/2}$ and 2p$_{1/2}$ orbitals of S$^{2-}$.[32] Quantitatively, the atomic composition obtained from the XPS data results in a S: Mo molar ratio of 2.27 ± 0.02 excluding the contribution of MnS residues. The small discrepancy to the stoichiometric ratio of MoS$_2$ can be attributed to defects at the nanocomposite's surfaces.[33] The overlapped peaks of C-C and C=O in the C1s scan and the strong peak in the O1s scan suggest the presence of oxygen containing groups in the carbon layer which contribute to the tight bonding between the carbon layer and the MoS$_2$ nanosheets. The presence of N1s peaks can be attributed to the PDA layers, because dopamine is the only raw material which contains the element N. The peaks at 398.5 and 400.9 eV correspond to pyridinic and graphitic N, respectively. [34] The N 1s scan in C@MoS$_2$@C exhibits the same peak position but a different peak ratio as compared with the PDA-derived carbon hollow sphere sample (cf. Fig. S6). The different peak ratio in the C@MoS$_2$@C sample may be attributed to the interaction between the inner PDA layer and the Mo precursors since the interaction will change the chemical environment of N. [26] This experiment hence offers evidence that PDA is preserved after the hydrothermal treatment.

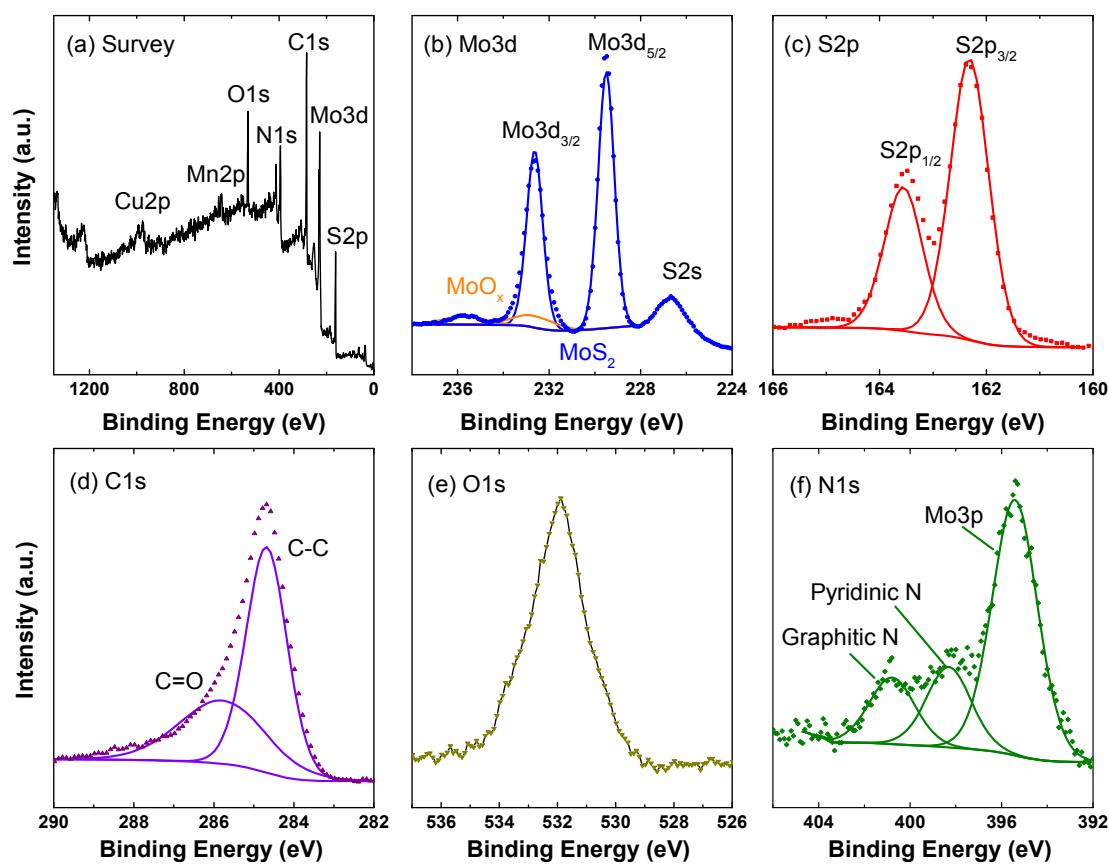

Fig. 5 XPS measurement on C@MoS$_2$@C: (a) survey scan, (b)~(f) high resolution scan of Mo3d, S2p, C1s, O1s, and N1s.

In order to investigate the Li$^+$ storage properties of the C@MoS$_2$@C hollow spheres, selected cycles of the cyclic voltammetry (CV) sweeps for both the sandwiched hollow spheres and the pure MoS$_2$ assembly have been examined. Typical redox features of the MoS$_2$ system are observed in the CV of C@MoS$_2$@C (Fig. 6b of the SI), which exhibits two distinct reduction peaks at 1.1 V and 0.6 V in the first cathodic scan. The former is attributed to the intercalation of Li$^+$ between the MoS$_2$ layers to form Li$_x$MoS$_2$, while the latter is assigned to the conversion from Li$_x$MoS$_2$ to metallic Mo and Li$_2$S.[16, 35] These reduction peaks weaken in the 2$^{nd}$ sweep and disappear in subsequent cycles because the resultant Mo nanoparticles are embedded in a Li$_2$S matrix during the conversion reaction and do not react back to MoS$_2$. Instead, Li$_2$S and S form a reversible redox couple which is indicated by the reduction/oxidation peaks at around 1.7-2.0/2.3 V.[36] From the CV data, one can conclude that the sandwiched hollow spheres exhibit better electrochemical activity and stability than the pure MoS$_2$ assembly. In particular, additional redox peaks around 1.2-1.7 V/1.2-2.0 V occur from the 2$^{nd}$ cycle on in the case of the C@MoS$_2$@C hollow spheres, which may relate to the lithiation-delithiation of the amorphous Mo/Li$_2$S matrix or amorphous MoS$_x$.[37] Furthermore, the oxidation peak at around 2.3 V of the pure MoS$_2$ assembly shifts to higher voltages with ongoing cycling, corresponding to a higher energy barrier during the lithiation-delithiation process.

The electrochemical performance was further investigated by galvanostatic cycling with potential limitation (GCPL) and electrochemical impedance spectroscopy (EIS). Fig. 6a shows charge-discharge profiles for the 1$^{st}$, 2$^{nd}$, 5$^{th}$, 10$^{th}$, and 50$^{th}$ cycle of the C@MoS$_2$@C hollow spheres at a C-rate of 0.1 C (1 C = 670 mA g$^{-1}$ for MoS$_2$ based on the conversion reaction). There are three potential plateaus at 1.7 V, 1.2 V and 0.6 V in the initial discharge process which is in good agreement with the reduction peaks of the first cathodic CV scan (Fig. S10 of the SI). The first plateau comes from the Li$^+$ intercalation into MnS impurities,[38] while the second and third plateaus are attributed to the intercalation process and conversion reactions of MoS$_2$, respectively. The initial discharge capacity of the sandwich hollow spheres is 1372.6 mA h g$^{-1}$, remaining 871.9 mA h g$^{-1}$ after 50 cycles. The charge/discharge voltage profiles do not show obvious changes from the 5$^{th}$ cycle on, revealing the good cycling stability of the electrode.

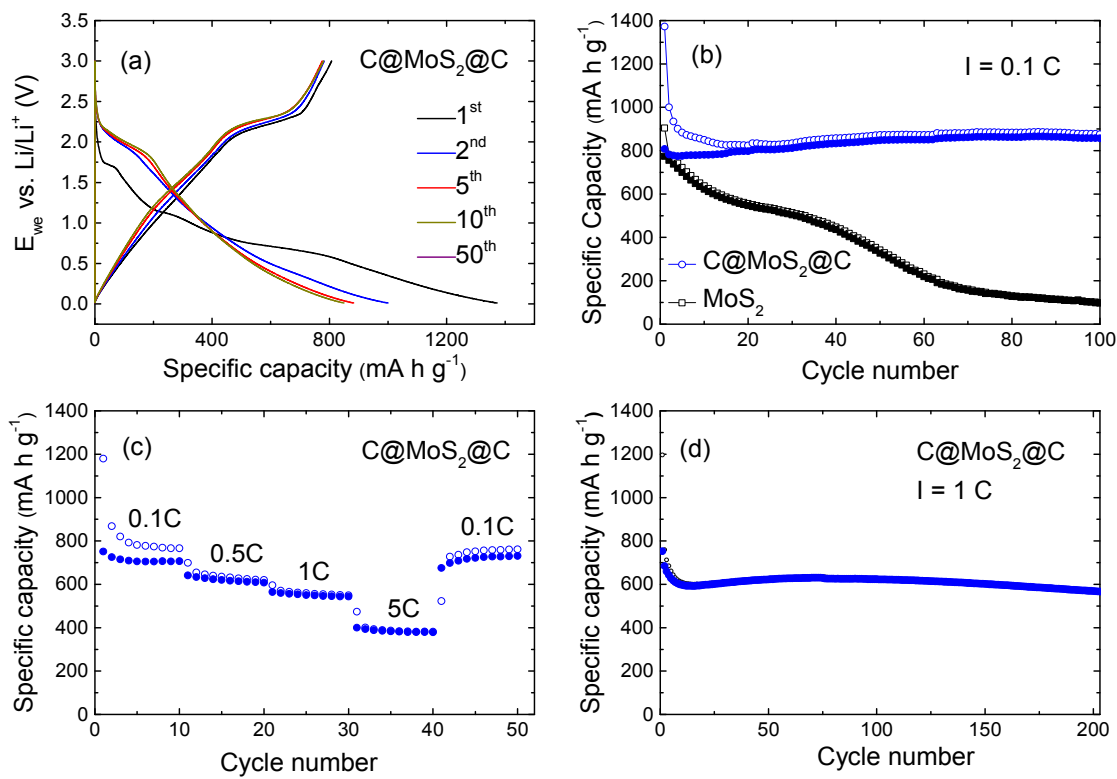

Fig. 6 (a) GCPL measurement of the C@MoS$_2$@C hollow spheres. (b) Cycling performance of pure MoS$_2$ assembly and C@MoS$_2$@C hollow spheres at 0.1 C. (c) Rate performance of the C@MoS$_2$@C hollow spheres. (d) Long term measurements of C@MoS$_2$@C hollow spheres at 1 C.

The good cycling stability of the hierarchically structured anode material can be observed directly from Fig. 6b and d. At the current density of 0.1 C(Fig. 6 (b)), the specific capacity of the C@MoS$_2$@C hollow spheres becomes stable after the first few cycles and reaches 856.7 mA h g$^{-1}$ at 100$^{th}$ cycle. The Coulombic efficiency is higher than 98% from cycle 15 onwards. In particular, the cycling stability of C@MoS$_2$@C clearly exceeds the one of the pure MoS$_2$ assembly. While the capacities of the hollow spheres stay more or less constant or even increase after considerable irreversible losses in the first ~15 cycles, the pure MoS$_2$ assembly shows continuous capacity fading of around 1% per cycle. Actually, the stable performance of the electrode can be extend to 200 cycles as shown in the long term GCPL measurement of C@MoS$_2$@C at 1 C (Fig. 6 (d)). After the stabilization process within the first 15 cycles, the discharge capacity reaches 600 mA h

g$^{-1}$ and remains 571 mA h g$^{-1}$ after 200 cycles, exhibiting only a decay of 0.026% per cycle (see Tab. S1). This value is smaller than most of the MoS$_2$ based anode materials reported in literature.[16, 37, 39]

The structural benefits of the hollow sandwich spheres also provide the electrode with excellent rate performance. The corresponding change of charge/discharge capacities with different current rates is shown in Fig. 6 (c). The initial discharge capacity at 0.1 C is 1180.2 mA h g$^{-1}$ and stabilizes around 800 mA h g$^{-1}$ from the second cycle on. This is consistent with the GCPL data of Fig. 6 (b). Subsequently, the discharge capacity decreases to 640 mA h g$^{-1}$, 560 mA h g$^{-1}$, and 382 mA h g$^{-1}$ when the current rate is increased to 0.5 C, 1 C and 5 C, respectively. However, the discharge capacity changes back to almost the initial value (763 mA h g$^{-1}$) when the current rate returns to 0.1 C. The data show that increase in current does not lead to huge capacity fading, demonstrating the superior rate performance of the electrodes derived from the stable hierarchical hollow structures.

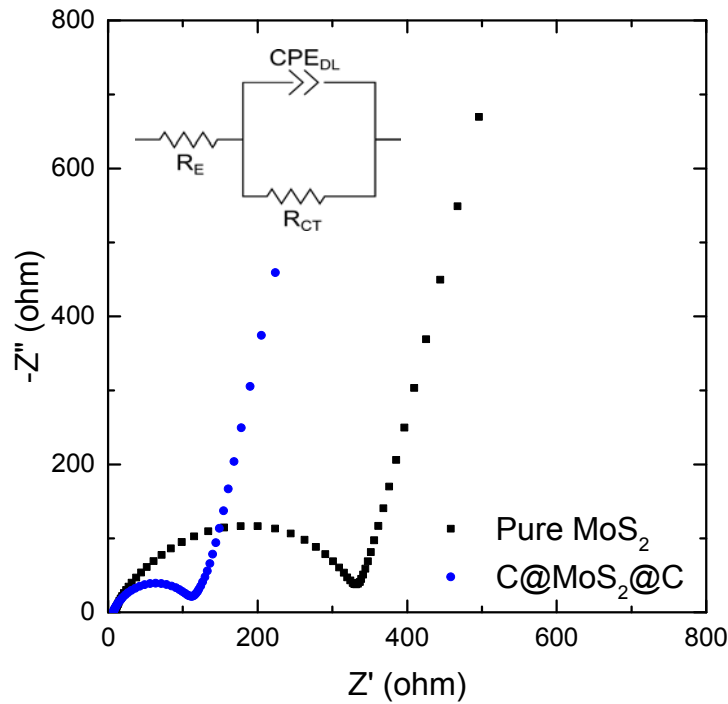

Fig. 7 Impedance measurements of the C@MoS$_2$@C hollow spheres and pure MoS$_2$ assembly before cycling. The inset is the equivalent circuit.

Electrochemical impedance spectra of pure MoS$_2$ assembly and C@MoS$_2$@C hollow spheres obtained at frequencies between 100 kHz and 0.1 Hz provide further insight into the electrochemical processes. The Nyquist plots of both samples before cycling, as shown in Fig. 7, exhibit depressed semi-circles in the high frequency range and a slope-like behavior at low frequencies corresponding to the charge transfer resistance between electrolyte and electrode material, and Li$^+$ diffusion impedance respectively. The semi-circles are described by means of a generalized RC-circuit with electrolyte resistance $R_E$, charge transfer resistance $R_{CT}$, and a constant phase element $CPE_{DL}$ for the electrical double layer, using the *Z Fit* function of the *EC-Lab* software (Bio-Logic). The used equivalent circuit is shown in the inset of Fig. 7, and the calculated parameters are listed in Tab. S2. It is found that the $R_{CT}$ of C@MoS$_2$@C hollow spheres is only 1/3 the value of the pure MoS$_2$ assembly, demonstrating the remarkable promotion of charge transfer in the hierarchically structured electrode. The reduced resistance can be attributed to the increased conductivity, and shortened and more efficient electron transfer pathways, originating from the carbon sandwiching of the MoS$_2$ nanosheets. The charge transfer resistances of the two samples decreases sharply after cycling as a result of the formation of SEI layers during the first few dis/charging processes (Fig. S11 of the SI). The low $R_{CT}$ of C@MoS$_2$@C hollow spheres remains constant after 20 cycles, and shows a slight increase after 50 cycles; however, the $R_{CT}$ of the pure MoS$_2$ assembly electrode increases strictly especially after 20 cycles as a result of structural destruction of the active material during cycling. In contrast, the sandwich hollow spheres exhibit a faster charge transfer rate and a highly stable nanostructure, which yield the high long-term cycling stability and good rate performance presented above.

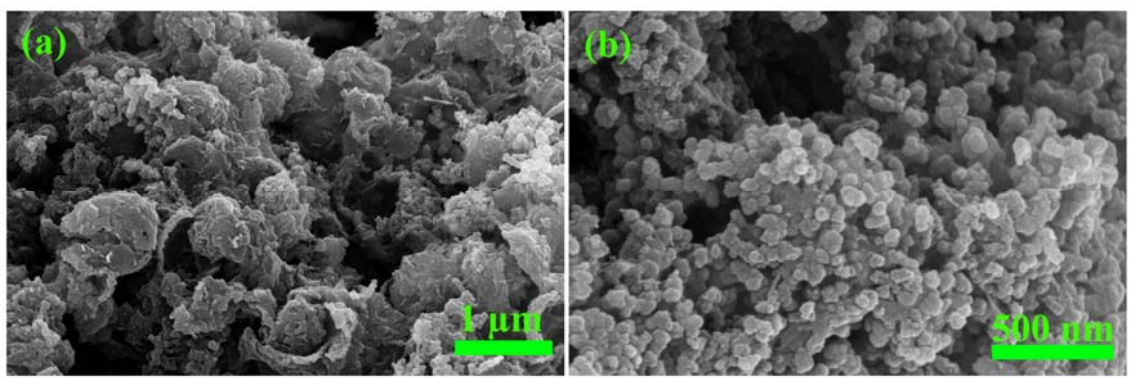

Fig. 8 SEM images of the samples after 50 cycles at 0.1 C: (a) C@MoS$_2$@C hollow spheres; (b) pure MoS$_2$ assembly.

The results presented above clearly demonstrate the structural superiority of the C@MoS$_2$@C hollow spheres. Firstly, the hollow interior and carbon layers of the material play a crucial role in buffering the mechanical stress induced by the volumetric expansion during the charge-discharge process. Secondly, the hierarchical structure as well as the interaction between the inner carbon layer and MoS$_2$ prevents the MoS$_2$ nanosheets from agglomeration during the synthesis process. Thirdly, the sandwich structure also prevents the active material (i.e. sulphur) from being dissolved in the electrolyte. At last, the double carbon layers scaffold is stable enough to stabilize the hollow structure, which is proven by the SEM images of the hollow sphere sample after 50 cycles at 0.1 C in Fig. 8a. In contrast, the pure MoS$_2$ assembly transforms to irregular shaped nanoparticles after the same cycling procedure (Fig. 8b). In fact, the as-prepared C@MoS$_2$@C hollow spheres show enhanced electrochemical properties even when compared with recent literature studies (see Tab. S3).[2, 7, 8, 13, 19]

**Conclusions**

In summary, we have rationally designed a modified template method for synthesizing hierarchical C@MoS$_2$@C hollow spheres. As an anode material of LIBs, the as-prepared sample exhibits high initial discharge capacity of 1372.6 mA h g$^{-1}$ at 0.1 C, quite stable cycling performance (0.026% fading per cycle at 1 C within 200 cycles) and prominent rate performance. This enhanced electrochemical performance benefits from both the advantages of the hierarchical sandwich hollow structure and the outstanding properties of the components. Due to the structural and componental advantages, hollow sandwich spheres are also promising in electrocatalic[40] applications. The reported synthesis route offer a facile and universal way to design and produce other functional nanomaterials.

**Experimental Methods**

MnCO$_3$ spheres were fabricated by using a modified co-precipitation method according to the literature.[41] MnSO$_4$·H$_2$O and NaHCO$_3$ powder were first dissolved in distilled water with the concentration of 0.04 mol L$^{-1}$ and 0.4 mol L$^{-1}$ respectively. 10.0 % v/v ethanol were then added to the above solutions under vigorous stirring. After completely dispersion of the solutions, NaHCO$_3$ solution were added to the MnSO$_4$ solution and kept stirring for 3 h. The milky white precipitate were collected after centrifuge, washing and drying. (NH$_4$)$_2$MoS$_4$ solution was obtained by reacting 200 mg ammonium molybdate tetrahydrate (AHM, Sigma-Aldrich, 81.0–83.0% MoO$_3$ basis), 3 mL ethylenediamine, 3 mL CS$_2$ and 50 mL distilled water at 85 °C overnight. The as-prepared MnCO$_3$ spheres were dispersed in 2 mg mL$^{-1}$ dopamine/Tris buffer solution (10 mM, pH=8.5) and stirred for 24 h to get a DOPA layer on the surface of MnCO$_3$ (DOPA@MnCO$_3$). After centrifuge, DOPA@MnCO$_3$ was dispersed into 30 mL distilled water before adding into 10 mL (NH$_4$)$_2$MoS$_4$ solution. The mixture was stirred for another 30 min and transferred into a 50 mL Teflon autoclave for hydrothermal synthesis afterwards. The autoclave was heated to 200 °C for 24 h. Then, the black precipitate was thoroughly washed with distilled water and treated with 2 M HCl solution for 24 h followed by 2 mg mL$^{-1}$ dopamine/Tirs solution for 24h. The final product was sintered at 900 °C for 10h under Ar atmosphere. For comparison, the MoS$_2$/MnS hybrid cubics synthesized by using MnCO$_3$ nanospheres as template in the same hydrothermal condition but without PDA protection layer. The pure MoS$_2$ assembly was produced from the same (NH$_4$)$_2$MoS$_4$ solution via the same hydrothermal condition without adding neither template nor the protection layer. Same sintering process was also applied after the hydrothermal treatment.

X-Ray powder diffraction (XRD) was performed in Bragg-Brentano geometry (Bruker-AXS D8 ADVANCE ECO) applying Cu-K$_{\alpha 1}$ radiation ($\lambda$ = 1.54056 Å). The step size $\Delta 2\theta$ was 0.02°. The morphology and microstructure of the sample was studied by scanning electron microscope (SEM, ZEISS Leo 1530) and transmission electron microscopy (TEM, JEM 2100F). X-ray photoemission spectroscopy (XPS) was carried out in an ESCALAB 250Xi ultra-high vacuum system using an Al Kα radiation source ($E$ = 1486.6 eV), a 900 μm spot size, and 20 eV pass energy. The samples were prepared on a copper plate with 10 mm in diameter. Three spots were measured on each sample. Thermogravimetric analysis was taken by using a TGA/DSC1 STARe System (Mettler Toledo) at a heating rate of 10 °C/min in air. The nitrogen physisorption measurements were performed at 77 K with a Gemini V (Micro-meritics, Norcross, GA) after degassing the sample at 120 °C for 2 h. Brunauer−Emmett−Teller (BET) analysis from the amount of $N_2$ absorbed at various relative vapor pressures (six points $0.05 < p/p_0 < 0.3$, nitrogen molecular cross-sectional area = 0.162 $nm^2$) was used to determine the surface area. The Barrett–Joyner–Halenda (BJH) analysis was applied to deduce the pore-size distribution.

Electrochemical studies were carried out using Swagelok-type cells.[42] Both the pure $MoS_2$ assembly and the $C@MoS_2@C$ hollow spheres electrodes were prepared from a mixture of the active material, carbon black (SuperP, Timcal) and polyvinylidene fluoride (PVDF, Sigma-Aldrich, 99%) binder with a weight ratio of 7:2:1, soaked in anhydrous 1-methyl-2-pyrrolidinone (NMP, Sigma-Aldrich, 99%). The slurry was pasted on a circular Cu plate (approx. 10 mm in diameter) with mass loading of about 1.0 mg $cm^{-2}$, dried overnight under vacuum at 80 °C and pressed. The resulting electrode was dried again in a vacuum oven at 80 °C for 4 h, and transferred into an Ar atmosphere glove box. The two-electrode Swagelok-type cell was assembled in the glove box using lithium foil as counter electrode and 1 M $LiPF_6$ in a 1:1 mixture of ethylene carbonate and dimethyl carbonate as the liquid electrolyte (Merck LP30). Cyclic voltammetry and galvanostatic cycling of the cells were performed at 25 °C between 0.01 and 3.0 V versus $Li^+/Li$ at various scan/current rates using a VMP3 multichannel potentiostat (Bio-Logic SAS). Electrochemical impedance spectroscopy (EIS) was carried out also by using the VMP3 multichannel potentiostat in the frequency range of 100 kHz to 0.1 Hz.

**Acknowledgements**

The authors thank I. Glass and Dr. Jan Freudenberg for experimental support. Financial support by the CleanTech-Initiative of the Baden-Württemberg-Stiftung (Project CT3 Nanostorage) and by the IMPRS-QD is gratefully acknowledged. Z.L. acknowledges financial support by the Chinese Scholarship Council, the Excellence Initiative of the German Federal Government, and by the Götze foundation.